\documentclass[sigconf]{acmart}
\renewcommand\footnotetextcopyrightpermission[1]{} 
\usepackage{microtype}
\usepackage{graphicx}
\usepackage{booktabs}
\usepackage{subcaption}
\usepackage[utf8]{inputenc}
\usepackage[T1]{fontenc}
\usepackage{url}
\usepackage{amsfonts}
\usepackage{nicefrac}
\usepackage{xcolor}
\usepackage{caption}
\usepackage[ruled,vlined]{algorithm2e}
\usepackage{algorithmic}
\usepackage{amsmath}

\usepackage{bm}
\usepackage{multirow}
\usepackage{makecell}
\usepackage{tabularx}
\usepackage{footmisc}
\usepackage{listings}
\usepackage{pgfplots}
\usepackage{titlesec}
\usepackage{hyperref}
\usepackage{fancyhdr}

\copyrightyear{2022}
\acmYear{2022}
\setcopyright{acmcopyright}\acmConference[ICPP '22]{51st International Conference on Parallel Processing}{August 29-September 1, 2022}{Bordeaux, France}
\acmBooktitle{51st International Conference on Parallel Processing (ICPP '22), August 29-September 1, 2022, Bordeaux, France}
\acmPrice{15.00}
\acmDOI{10.1145/3545008.3545087}
\acmISBN{978-1-4503-9733-9/22/08}

\begin{document}

\title{Tesseract: Parallelize the Tensor Parallelism Efficiently}

\author{Boxiang Wang}
\email{boxiangw@hpcaitech.com}
\affiliation{%
  \institution{HPC-AI Technology Inc.}
  \city{Singapore}
  \country{Singapore}
}

\author{Qifan Xu}
\email{QifanXu@mednet.ucla.edu}
\affiliation{%
  \institution{University of California, Los Angeles}
  \city{Los Angeles}
  \country{United States of America}
}

\author{Zhengda Bian}
\email{bian.zhengda@hpcaitech.com}
\affiliation{%
  \institution{HPC-AI Technology Inc.}
  \city{Beijing}
  \country{China}
}

\author{Yang You}
\email{youy@comp.nus.edu.sg}
\affiliation{%
  \institution{National University of Singapore}
  \city{Singapore}
  \country{Singapore}
}

\renewcommand{\shortauthors}{Anonymous Author, et al.}

\begin{abstract}
Together with the improvements in state-of-the-art accuracies of various tasks, deep learning models are getting significantly larger. However, it is extremely difficult to implement these large models because limited GPU memory makes it impossible to fit large models into a single GPU or even a GPU server. 
Besides, it is highly necessary to reduce the training time for large models. 
Previous methods like Megatron-LM implemented a 1-Dimensional distributed method to use GPUs to speed up the training. 
However, these methods have a high communication overhead and a low scaling efficiency on large-scale clusters.
To solve these problems, we propose Tesseract, highly scalable tensor parallelism with a novel design. It increases efficiency by reducing communication overhead and lowers the memory required for each GPU. 
By introducing the novel dimension into tensor parallelism, Tesseract greatly increases the memory capacity of tensor parallelism. 
Concretely, this new dimension furthermore increases the degree of tensor parallelism.
Compared to previous 1-D and 2-D methods, Tesseract manages to reduce the communication cost on each layer, resulting in speedups of 1.38x and 1.53x respectively with strong scaling. 
In weak scaling experiments, Tesseract achieves a maximum of 4.0/1.7 times inference speedup and 3.4/1.7 times throughput improvement compared to 1-D/2-D methods, respectively.
By introducing Tesseract, we offer a more efficient and scalable way to implement large deep learning models with limited GPU resources.

\end{abstract}


\maketitle
\pagestyle{plain}

\section{Introduction}

The size of Neural Network models \cite{gpt1, gpt2, gpt-3, BERT, Transformer} is growing at a very high speed. 
However, the processor's performance does not have a similar improvement.
Thus, we need to scale the training to multiple processors or even multiple servers. 

To alleviate the poor generalization performance resulting from large batches, researchers proposed efficient optimizers like LAMB \cite{you2020large} and LARS \cite{you2017large}.

Data parallelism does a good job of speeding up the training \cite{goyal2017accurate, akiba2017extremely, jia2018highly, you2018imagenet, ying2018image, kumar2019scale, yamazaki2019yet}. 
However, in a situation where the memory of a single device can not host the whole model, only using data parallelism would lead to a failure. 
Activation checkpointing \cite{activations} alleviated the memory constraints by recomputing. 
Mixed precision training \cite{mixed} reduced both the memory overhead and computation cost. 
ZeRO-infinity \cite{zeroinfinity} used CPU memory and NVMe to extend the GPU memory while achieving decent throughput by largely overlapping computation and communication. 
Yet another direction is model parallelism. 
GPipe \cite{gpipe} and PipeDream \cite{pipedream} split the model vertically. 
Each node hosts a partition of the whole model, takes input from the previous node, and sends the output to the next node.

These techniques are orthogonal to our method in this paper.
Another option is to partition the model by the operator, or horizontally. 
Megatron-LM \cite{megatron} introduced a 1-Dimensional distributed method to speed up the training process of huge models.
Optimus \cite{optimus} leveraged SUMMA \cite{SUMMA} and took a step further to improve both the memory and communication efficiency. 
By using these methods, language models with a growing amount of parameters could be trained on separated processors whose memory could not afford the training procedure independently. 
However, due to the property of SUMMA, the communication within a $[q,q]$ shape processor array takes up to $2q^3$ times of the communication among its column and row. 
Together with the increment of $q$, the increment of communication between GPUs will also lead to a reduction of SUMMA's efficiency. 
Due to this consideration, we design and implement a novel method that could significantly reduce the communication overhead. 
From our experiments, we find that with the same number of processors and input tensors, our method would have less communication between GPUs on each layer and thus increase overall efficiency.

2.5D Matrix Multiplication method \cite{mm25d} was proposed to improve the efficiency of Cannon's algorithm \cite{cannon1969cellular}.
Yet with many restrictions of Cannon's Algorithm and a huge amount of shift operations, both the 2.5D Matrix Multiplication method and Cannon's Algorithm are not suitable, thus we need to invent a novel method to enhance the performance. According to our calculation, with a total amount of 64 processors, the communication needed for Cannon's Algorithm is 31.5 times the communication needed for Tesseract, and the communication needed for the 2.5D algorithm is 3.75 times the communication needed for Tesseract.
Inspired by both SUMMA and 2.5-Dimensional Matrix Multiplication, we introduce Tesseract for language models to overcome the abundance of unnecessary communication resulting from the increasing size of language models and increase the efficiency of tensor parallelism.

In this paper, we implement Transformer with Tesseract to examine its performance. In order to make Tesseract satisfy the demands of huge models, we develop the parallelization scheme for matrix-multiplication sections and non-matrix-multiplication sections separately. The transformer contains two main matrix-multiplication sections, a feed forward section, and a multi-head attention section.
For feed forward layers, we apply Tesseract algorithms to get the output and then store the parameter matrices inside each processor for the next computation to avoid waste of communication.
To implement the distributed method for multi-head attention layers, we compute corresponding $Q, K, V$ matrices and then attention output respectively. 
The attention would be computed separately on each processor, and to obtain $Q, K, V$ matrices the algorithm will perform matrix multiplication among the whole layer. 
For non-matrix-multiplication sections like layer normalization, the algorithm will broadcast the matrix along the column, allowing the distributed processors to work concurrently. 
Tesseract manages to reduce the communication cost on each layer, which could get a 2.1 times speedup according to our weak scaling result between 3-D $[4,4,4]$ arrangement and 2-D $[8,8,1]$ arrangement. 
Compared to other parallelization structures, Tesseract reached a 1.4x speedup and a 1.5 times speedup accordingly compared to Megatron-LM \cite{megatron} and Optimus \cite{optimus} in a strong scaling setting, which proves the efficiency of our parallelization strategy.
Meanwhile, with a weak scaling setting, Tesseract achieves a maximum of 4.0/1.7 times inference speedup and 3.4/1.7 times throughput improvement compared to 1-D/2-D methods, respectively.
Also, due to the price and build issues, GPUs are not always in the arrangement people wanted, Tesseract offers a flexible depth and dimension which could help users use their GPUs in the most efficient way.
\section{Preliminary and related work}
\label{section:preliminary}

\subsection{Cannon's Algorithm}
\begin{figure*}[h]
  \centering
  \setbox1=\hbox{\includegraphics{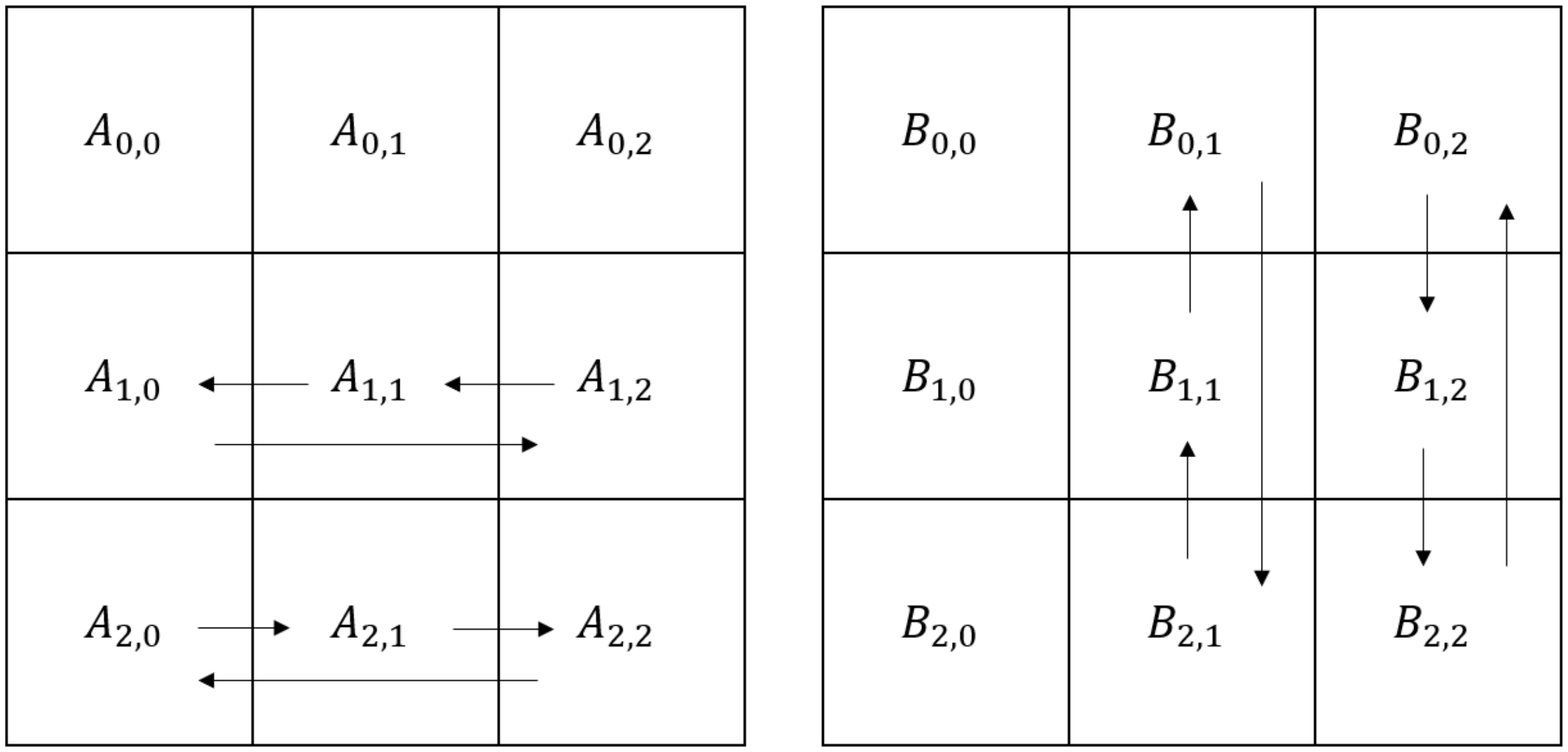}}
  \subcaptionbox{Initialization for matrix $A$ and matrix $B$}{\centering\includegraphics[width=85mm,scale=0.25]{figure/init.pdf}}
  \qquad
  \setbox1=\hbox{\includegraphics{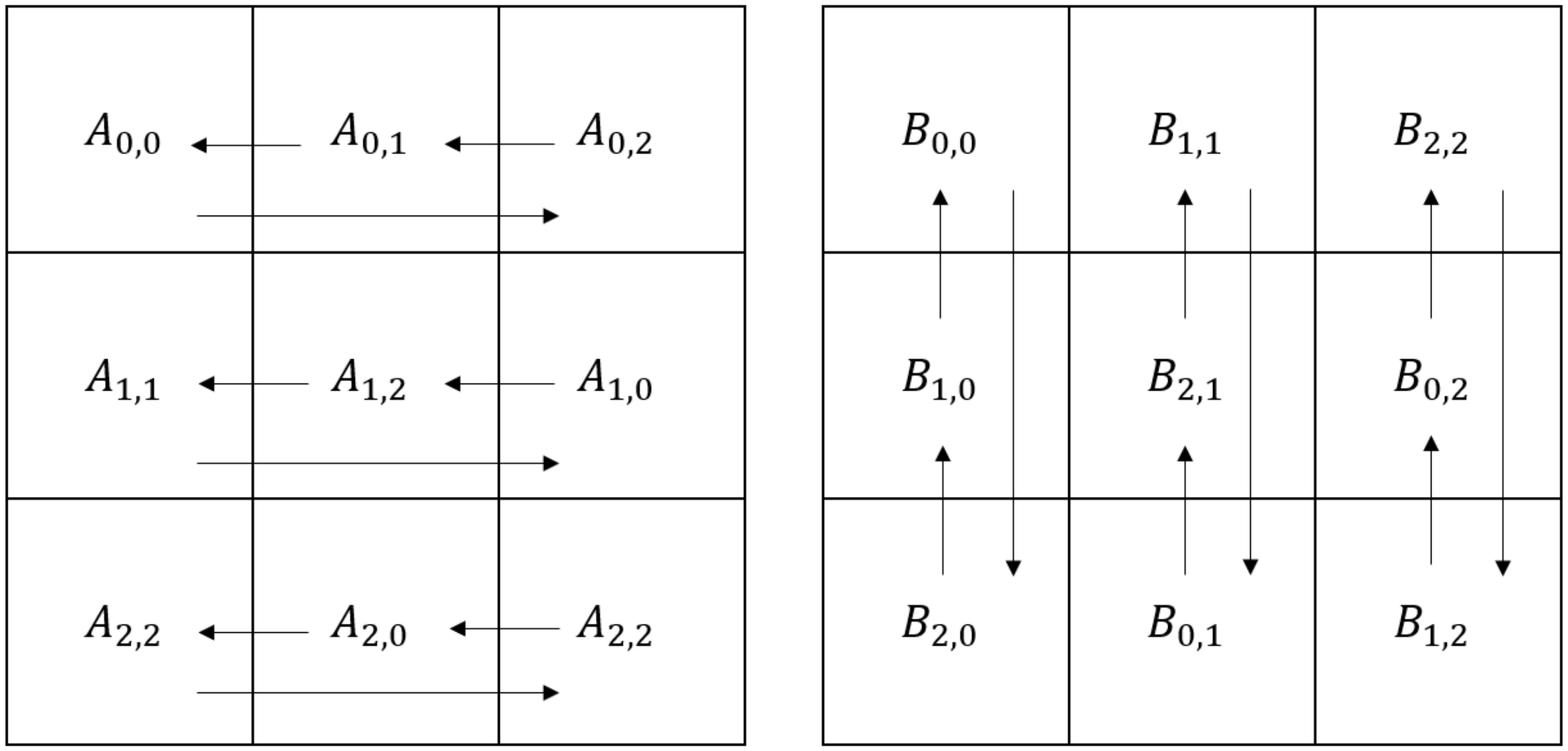}}
  \subcaptionbox{Shift of matrix $A$ and matrix $B$ after initialization }{\centering\includegraphics[width=85mm,scale=0.25]{figure/skew.pdf}}
  \caption{For Cannon's Algorithm, it requires a initialization where partition $A_{i,j}$ needs to shift left by $i$, partition $B_{i,j}$ needs to shift up by $j$, as shown in (a). After initialization, Cannon's Algorithm will compute locally, and shift all $A_{i,j}$ left by 1, all $B_{i,j}$ up by 1, as shown in (b). In this figure, we have $C_{i,j} = A_{i,0}B_{0,j} + A_{i,1}B_{1,j} + A_{i,2}B_{2,j}$}
  \label{fig:cannon's algo}
\end{figure*}

Cannon's Algorithm \cite{cannon1969cellular} introduced by Cannon in 1969 is widely used for matrix multiplication on distributed systems. It could be described as Figure \ref{fig:cannon's algo}. Cannon's algorithm applied the primary-and-secondary model and the divide-and-conquer model. Its zeroth processor will arrange the I/O for all other processors. It controls broadcast and reduce procedures. In another perspective, it divides the multiplication of two input matrices into small pieces, after the calculation on each processor, the final result will be the combination of results from all processors. Cannon's algorithm not only parallelizes the matrix multiplication but also reduces the storage needed for each processor. With number of arithmetic operations per process equals to $\frac{n^3}{p}$, memory size per process scales with $\Omega(\frac{n^2}{p})$, where the matrix size is $[n, n]$ and the processor amount is $p$. The lower bound on communication time and estimated lower bound of latency are: 
\begin{equation}
    W = \Omega(\frac{\mbox{number of arithmetic operations}}{\sqrt{\mbox{memory size}}}) = \Omega(\frac{n^2}{\sqrt{p}}).
\end{equation}
\begin{equation}
    S = \Omega(\frac{\mbox{number of arithmetic operations}}{(\mbox{memory size})^{3/2}}) = \Omega(\sqrt{p}).
\end{equation}
The idea of Cannon's algorithm is described in Algorithm \ref{cannon}. Assume there are $p = q^2$ processors in a $[q, q]$ shape: Firstly, shift matrix $A$'s $[q, q]$ partitioned matrices left by their corresponding row number. Secondly, shift matrix $B$'s $[q, q]$ partitioned matrices up by their corresponding column number. After these two steps, the algorithm will start to compute the matrices and get the corresponding $C_{ij}$. Then all partitioned matrices $A_{ij}$ will be shifted left by one, all partitioned matrices $A_{ij}$ will be shifted up by one, and add the calculated $C_{ij}$ to the previous result, repeat this procedure by $n$ times. Output $C$ by combining all $C_{ij}$ accordingly.

\begin{algorithm}[h]
   \caption{2-D matrix multiplication Cannon's Algorithm ($p$ processors in a $[q, q]$ shape)}
   \label{cannon}
\begin{algorithmic}
   \STATE {\bfseries Input:} Matrix $A$ with size $[a, b]$; Matrix $B$ with size $[b, c]$
   \STATE {\bfseries Output:} Matrix $C = A * B$ with size $[a, c]$
   \STATE split $A, B$ in to $p$ parts to match the processor shape\;
   \STATE store $A_{ij}, B_{ij}$ into $p_{ij}$ accordingly\;
   \STATE store $C_{ij} = 0$ into $p_{ij}$ accordingly\;
   \FOR{$i,j$ {\bfseries in} $\{0,...,q-1\}$}
   \STATE shift $A_{ij}$ to $p_{i(j-i)}$\;
   \STATE shift $B_{ij}$ to $p_{(i-j)j}$\;
   \FOR{$t = 0$ {\bfseries to} $q - 1$}
   \STATE $C_{ij} = C_{ij} + A_{ij}*B_{ij}$\;
   \STATE shift the submatrix $A$ owned by $p_{ij}$ to $p_{i(j-1)}$\;
   \STATE shift the submatrix $B$ owned by $p_{ij}$ to $p_{(i-1)j}$\;
   \ENDFOR
   \ENDFOR
   \STATE combine all $C_{ij}$ accordingly to $C$\;
   \STATE \Return $C$
\end{algorithmic}
\end{algorithm}

\subsection{SUMMA}
\label{section:2.2}
The Scalable Universal Matrix Multiplication Algorithm (SUMMA) \cite{SUMMA} provides a more effective and efficient algorithm for 2-D matrix multiplication. For SUMMA, Algorithm \ref{summa} is the pseudo-code for $C = A * B$. The corresponding differentiation for $A^\prime, B^\prime$ can be calculated as:
\begin{equation}
    A^\prime = C^\prime  B^{T}, B^\prime = A^T  C^\prime
\label{equation: summa}
\end{equation}

By arranging the $p$ processors into a $q * q$ mesh, the matrices $A$ and $B$ are also partitioned to $p$ parts accordingly. After the partitions of $A$ and $B$ are sent to the corresponding processors, the SUMMA algorithm allows each processor to calculate in parallel. At the end of the computation, the algorithm returns the resulting matrix $C$, distributed among the processors, in the same manner as $A$ and $B$ are partitioned.

\begin{algorithm}[h]
   \caption{2-D matrix multiplication SUMMA ($p$ processors in a $[q, q]$ shape)}
   \label{summa}
\begin{algorithmic}
   \STATE {\bfseries Input:} Matrix $A$ with size $[a, b]$; Matrix $B$ with size $[b, c]$
   \STATE {\bfseries Output:} Matrix $C = A * B$ with size $[a, c]$
   \STATE split $A, B$ in to $p$ parts to match the processor shape\;
   \STATE store $A_{ij}, B_{ij}$ into $p_{ij}$ accordingly\;
   \STATE store $C_{ij} = 0$ into $p_{ij}$ accordingly\;
   \FOR{$i,j$ {\bfseries in} $\{0,...,q-1\}$}
   %\STATE $C_{ij} = 0$\;
   \FOR{$t = 0$ {\bfseries to} $q - 1$}
   \STATE broadcast $A_{it}$ in $p_{it}$ to $p_{ij}$\;
   \STATE broadcast $B_{tj}$ in $p_{tj}$ to $p_{ij}$\;
   \STATE $C_{ij} = C_{ij} + A_{it}*B_{tj}$\;
   \ENDFOR
   \ENDFOR
   \STATE combine all $C_{ij}$ accordingly to $C$\;
   \STATE \Return $C$
\end{algorithmic}
\end{algorithm}

\subsection{2.5-Dimensional Matrix Multiplication}
\label{2.5}
In 2011, E Solomonik et al. introduced a 2.5-D matrix multiplication method \cite{mm25d} to reduce communication for Cannon's Algorithm. This method is named 2.5-D because it has special cases of both 2-D and 3-D matrix multiplication. It uses multiple processors with a shape of $q*q*d$ where $p$ represents the number of processors, $d$ represents the depth of the processor group, and $p$ represents width and length. Compared with 2-D Cannon's Algorithm \cite{cannon1969cellular} and PDGEMM by ScaLAPACK, the 2.5-D algorithm could speed up the calculation with less communication cost. The lower bound on communication time is
\begin{equation}
    W = \Omega(\frac{\mbox{number of arithmetic operations}}{\sqrt{\mbox{memory size}}}) = \Omega(\frac{n^2}{\sqrt{dp}})
\end{equation}

The estimated lower bound of latency is

\begin{equation}
    S = \Omega(\frac{\mbox{number of arithmetic operations}}{(\mbox{memory size})^{3/2}}) = \Omega(\frac{p^{1/2}}{d^{3/2}})
\end{equation}
Where number of arithmetic operations per process equals to $\frac{n^3}{p}$, memory size per process scales with $\Omega(\frac{dn^2}{p})$, where the matrix size is $[n, n]$.
In special cases like $d = 1$, the 2.5-D algorithm degenerates to Cannon's algorithm; when $d=p^{1/3}$, it becomes a 3-D algorithm.

\subsection{Transformer}
Transformer \cite{Transformer} was published by Google in 2017. Before that, natural language processing (NLP) was dominated by recurrent neural networks like LSTM. Due to the sequential nature of these models, it was difficult to train with large batch sizes. Transformer broke the serial dependency and allowed each hidden vector to attend to any preceding ones. In this manner, the training of Transformer models could be elegantly formulated into basic matrix-matrix multiplications, thus becoming scalable.

The original Transformer consists of an encoder and a decoder. Encoder and decoder are again composed of multi-head attention and feed forward layers. In Megatron-LM, the architecture is adapted in a manner that the whole model consists of multiple identical Transformer layers. Each Transformer layer consists of a self-attention module and a multi-layer perceptron (MLP) module. MLP is simply two linear layers with an activation function in the middle, the first projecting the hidden vector to a higher dimension while the second projecting it back to the original hidden size. The self-attention module first uses a linear layer to project the original hidden vector into queries ($Q$), keys ($K$), and values ($V$). Multi-head self-attention is calculated as:
\begin{equation}
    A=\mathrm{softmax}(\frac{QK^T}{\sqrt{d}})V.
\end{equation}
The $A$'s are then rearranged back to the hidden size and undergo another linear layer to produce the output of the self-attention module. This design makes the self-attention module and MLP each have 2 linear layers, facilitating the row-column partitioning.

\begin{figure*}[h]
    \centering
    \includegraphics[width=110mm]{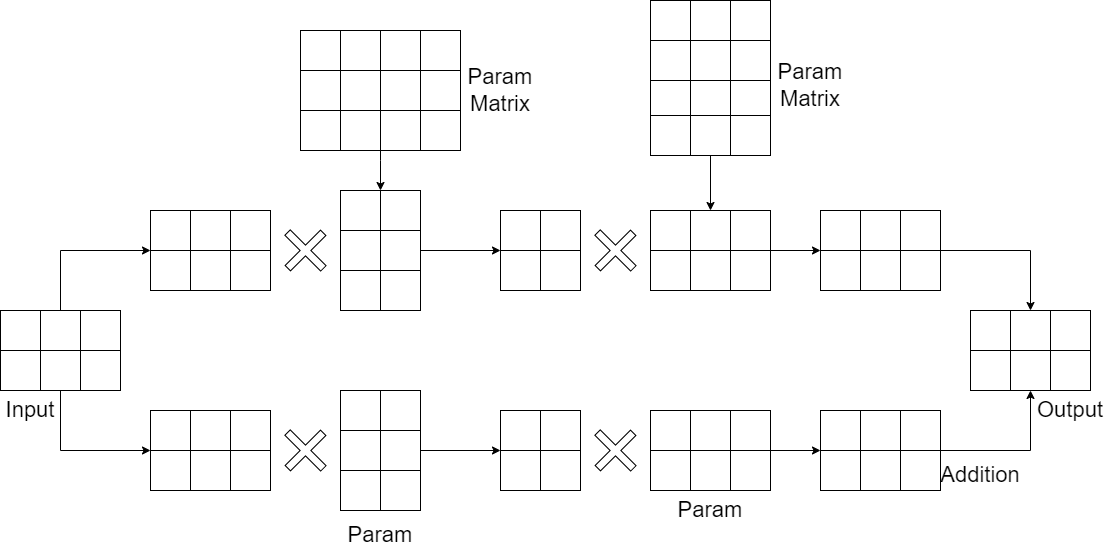}
    \caption{The tensor split method used in Megatron-LM, two different parameter matrices are split using different methods according to their columns and rows.}
    \label{fig:megatron}
\end{figure*}

\subsection{Megatron-LM}
Introduced by Nvidia in 2019, Megatron-LM \cite{megatron} proposed a tensor parallel structure for Transformer-like deep learning models we refer to as 1-D solution on tensor parallelism. In this particular setting, matrices involved in matrix multiplication will be split in one dimension and then conduct multiplication processes as shown in Figure \ref{fig:megatron}. As shown in the Figure, the input matrix with a shape of $[a, b]$ will conduct two different matrix multiplication in two different processes. During the matrix multiplication process, there are two parameter matrices with shapes $[b, 2c]$ and $[2c, b]$ respectively. If no tensor parallelism is involved in the process, the input matrix $[a, b]$ will conduct matrix multiplication with these two matrices accordingly and result in an output matrix with a shape of $[a, b]$. In the setting of Megatron, the two parameter matrices will be split according to their column and row accordingly, 
producing matrices with shapes of $[b, c]$ and $[c, b]$. The input matrix in two processes will conduct multiplication with the split parameter matrices and result in two different matrices with the shape of $[a, b]$. Finally, these two matrices will add together to get the final output matrix with the same shape as the input matrix $[a, b]$. As the first to introduce tensor parallelism, Megatron-LM could further speed up the process of model training for Transformer-like huge deep learning models.

\section{Tesseract}

\subsection{Tesseract}
\label{section:Tesseract}

With the high growth rate of parameter size in state-of-the-art neural network models and a much lower growth rate of fast memory, it is necessary to develop a tensor parallelism scheme to allocate the memory needed for neural network models into each processor. 
Under this circumstance, we introduce Tesseract, a novel tensor parallelism method, to reduce the communication overhead and memory cost.
Specifically, we design a novel tensor partitioning strategy with a unique arrangement of processors, it could be described as a highly scalable tensor parallelism method with a new dimension. 
To avoid misunderstanding, we define the following notations:

\begin{itemize}
  \setlength\itemsep{1em}
  \item Tesseract dimension: $q$
  \item Tesseract depth: $d$
  \item number of processors: $p$
  \item batch size: $b$
  \item hidden size: $h$
  \item sequence length: $s$
  \item number of Transformer layers: $N$
  \item memory: $M$
\end{itemize}

\begin{figure}[h]
    \centering
    \includegraphics[width=80mm]{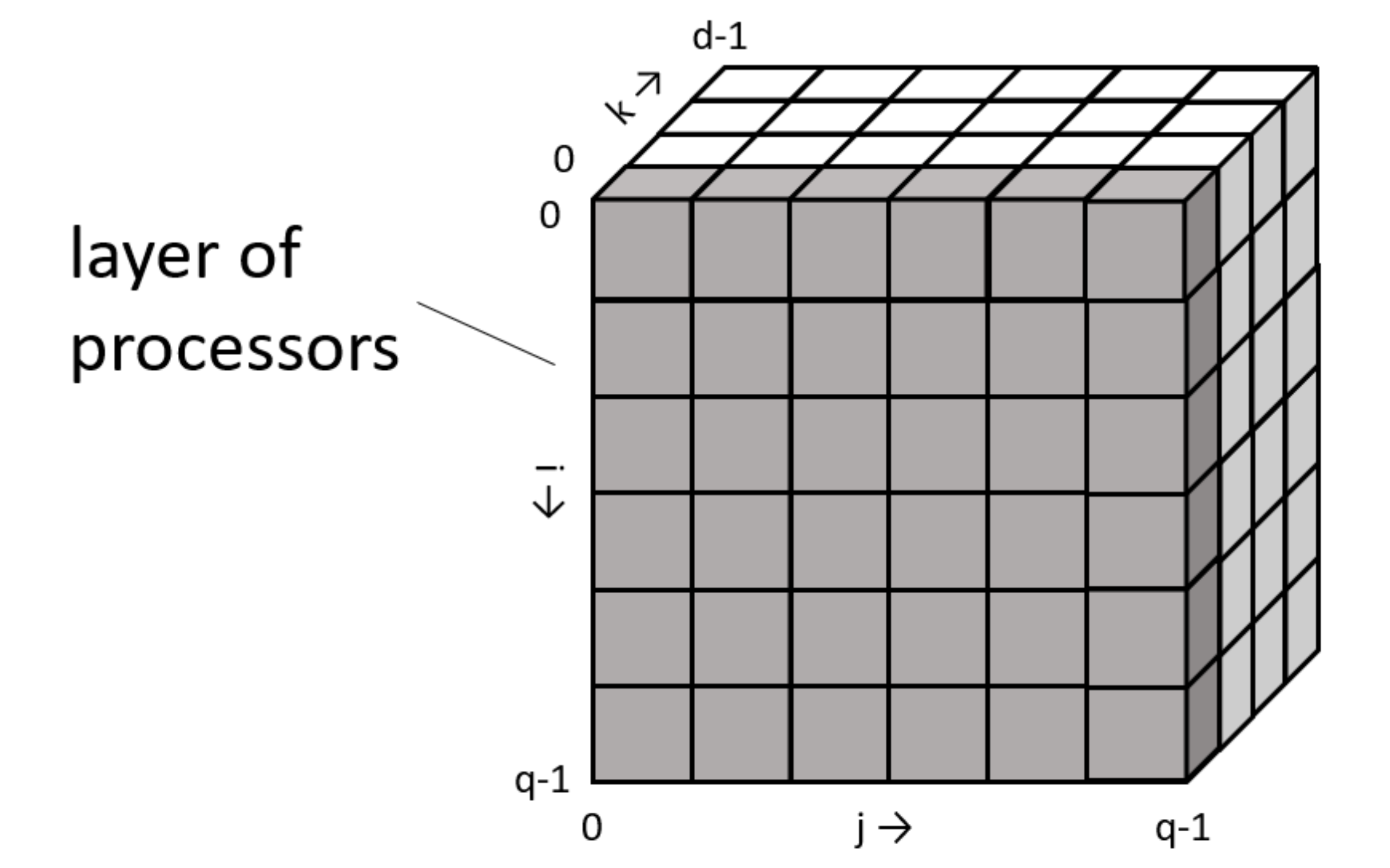}
    \caption{$p = dq^2$ processors in Tesseract arrangement of shape $[q, q, d]$}
    \label{fig:method}
\end{figure}

\begin{figure*}[!h]
  \centering
  \setbox1=\hbox{\includegraphics{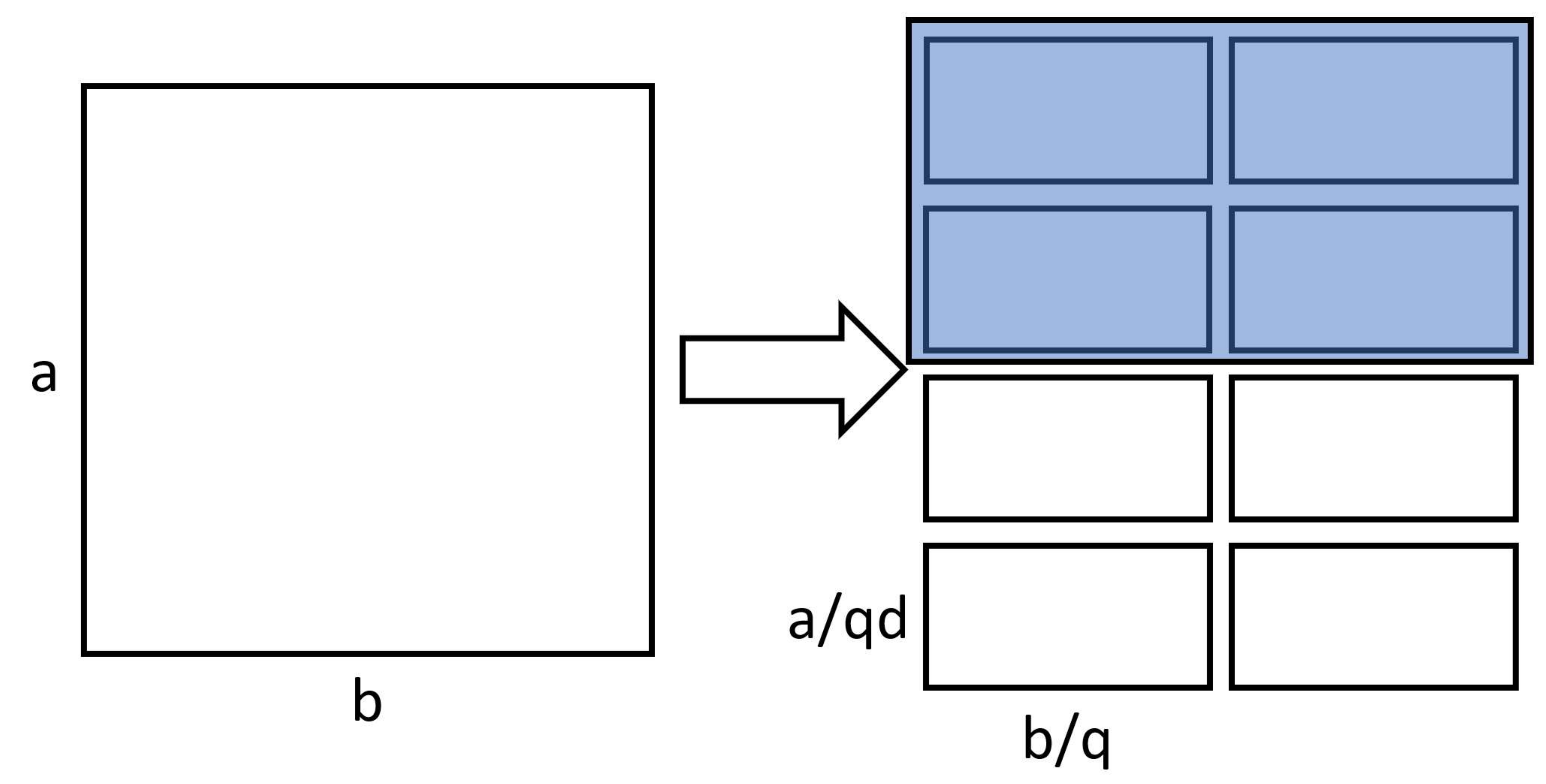}}
  \subcaptionbox{Partition of matrix $A$}{\centering\includegraphics[width=65mm,scale=0.25]{figure/A.pdf}}
  \qquad
  \setbox1=\hbox{\includegraphics{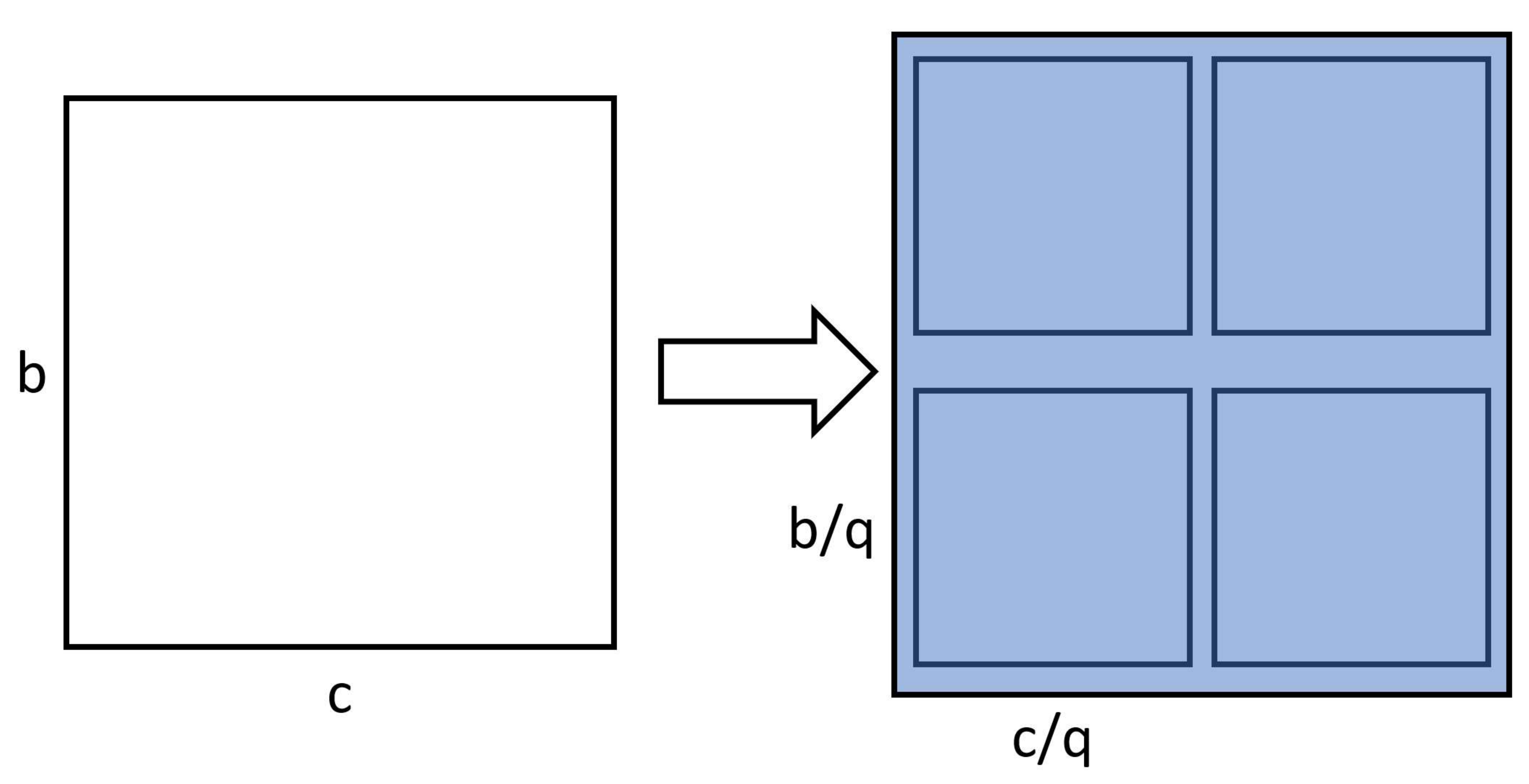}}
  \subcaptionbox{Partition of matrix $B$}{\centering\includegraphics[width=65mm,scale=0.25]{figure/B.pdf}}
  \qquad
  \setbox1=\hbox{\includegraphics{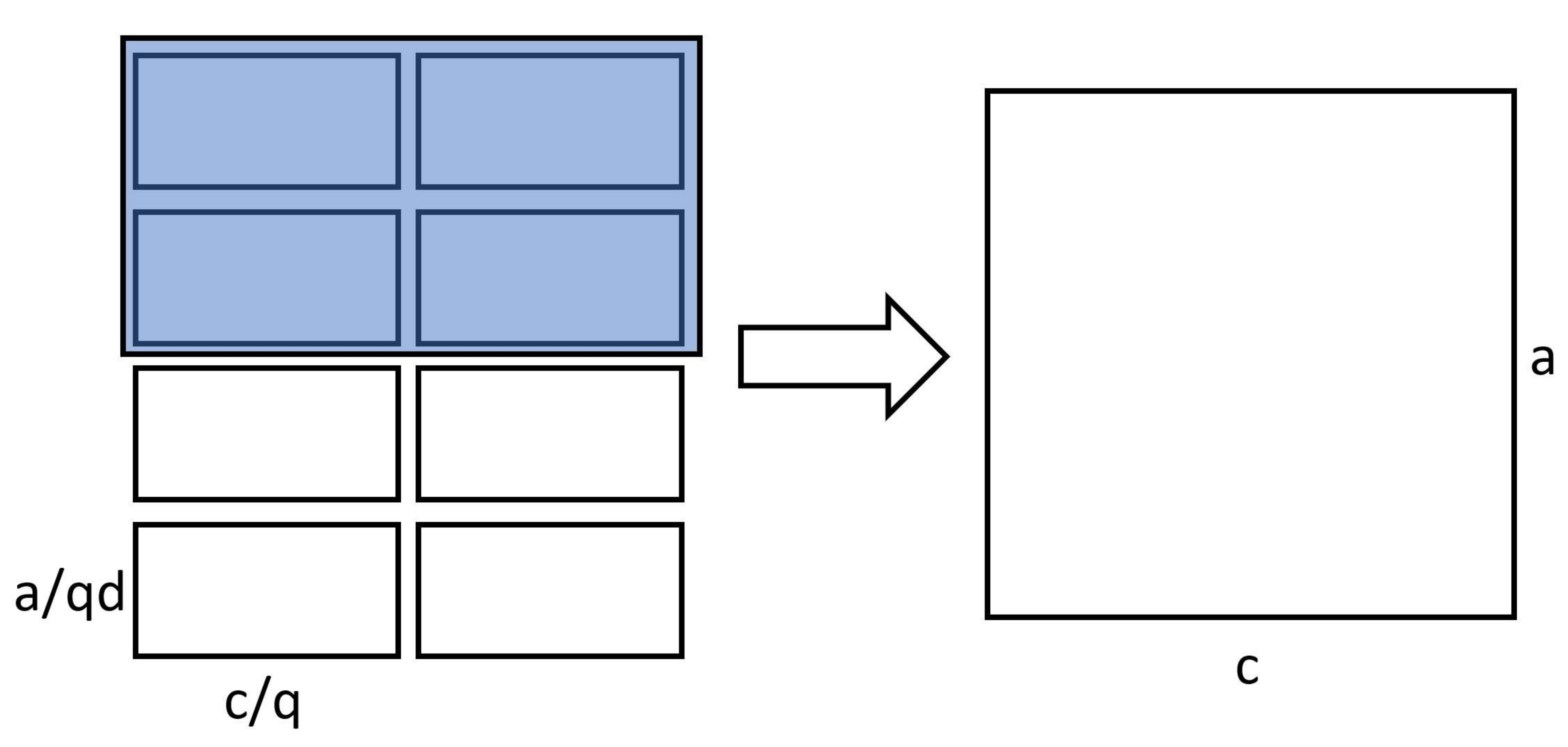}}
  \subcaptionbox{Combination of matrix $C$}{\centering\includegraphics[width=65mm,scale=0.25]{figure/C.pdf}}
  \caption{Method to split input matrices $A,B$, and method to combine output matrix $C$ assuming processors' shape $[q=2, q=2, d=2]$. The blue region representing a layer among the processors with a shape of $[q=2, q=2]$. (a) Matrix $A$ with shape $[a, b]$ will be split into $dq^2$ partitioned matrices with shape of $[a/qd, b/q]$, $[q,q]$ partitioned matrices will be stored in each layer. (b) Matrix $B$ with shape $[b, c]$ will be split into $q^2$ partitioned matrices with shape of $[b/q, c/q]$, $[q,q]$ partitioned matrices will be stored in each layer. (c) $dq^2$ partitioned matrices with shape of $[a/qd, b/q]$ will be combined into matrix $C$ with shape $[a, b]$.}
  \label{fig:split}
\end{figure*}

Where $p = dq^2$ and $1 \leqslant d \leqslant q$, $d = 1$ makes Tesseract a 2-D algorithm like SUMMA, and $d = q$ makes Tesseract a 3-D algorithm. The $p$ processors will be arranged in a shape of $[q, q, d]$ as shown in Figure \ref{fig:method}. Tesseract splits input matrix $A$ with a shape of $[a, b]$ and matrix $B$ with a shape of $[b, c]$ into partitions to match the arrangement of $p$ processors. After the calculation, Tesseract outputs the matrix $C$ with a shape of $[a, c]$ combined from all $C_{ijk}$ on different processors. The procedure of our matrix multiplication is described in Algorithm \ref{Tesseract algo}. The method of how matrices will be split and combined is shown in Figure \ref{fig:split}. After calculation of the partitioned matrix multiplication between size $[a/qd, b/q]$ and $[b/q, c/q]$, the respective result matrices will be in shape $[a/qd, c/q]$ which are stored in respective processors just like input matrix $A$, thus the algorithm could use broadcast and reduce in the same way for matrices $A$ and $C$. 

For calculation of $C = A * B^T$, we use a algorithm which broadcasts $B_{tjk}$ within its column and computes $C_{ijk} = A_{ijk} * B_{tjk}^T$, then the algorithm reduces the $C_{ijk}$ within its column to $C_{itk}$. Similar for function $C = A^T * B$, the algorithm broadcasts $B_{itk}$ within its row, computes $C_{ijk} = A_{itk}^T * B_{tjk}$, then reduces the $C_{ijk}$ within its column to $C_{tjk}$.

For computation of matrices' gradients, the function \ref{equation: summa} in section \ref{section:2.2} is applied in Tesseract. For matrix $A, C$, the $dq^2$ partitioned matrices will return $dq^2$ partitioned gradient matrices, but for matrix $B$, the $q^2$ partitioned matrices will return $dq^2$ partitioned gradient matrices, in order to get a correct shape of gradients, our algorithm applied $all\_reduce$ function after the computation of $B^\prime$ on processors with same row and column but different depth.

According to Figure \ref{fig:split}, each processor stores a partition of matrix $A, B, C$, and their respective sizes are $[\frac{a}{d*q}, \frac{b}{q}], [\frac{b}{q}, \frac{c}{q}], [\frac{a}{d*q}, \frac{c}{q}]$. Thus we could compute the memory needed for each processor to perform a single matrix multiplication operation:
\begin{align}
    M_{Tesseract} &= \frac{a}{d*q}*\frac{b}{q}+\frac{b}{q}*\frac{c}{q}+\frac{a}{d*q}*\frac{c}{q}\\       
                  &= \frac{a*b}{p}+\frac{b*c*d}{p}+\frac{a*c}{p}
\end{align}
Compared to Megatron-LM, it operates matrix multiplication with the size of matrices in $[a, b], [b, \frac{c}{p}], [a, \frac{c}{p}]$, and the memory required for each processor is:
\begin{align}
    M_{Megatron-LM} &= a*b+b*\frac{c}{p}+a*\frac{c}{p}\\       
                    &= a*b+\frac{b*c}{p}+\frac{a*c}{p}
\end{align}
The comparison between $M_{Tesseract}$ and $M_{Megatron-LM}$ is clear that same memory is needed for matrix $C$, and Megatron-LM requires $p$ times more memory to store matrix $A$. Although Tesseract spends more memory on matrix $B$, it is negligible since $p = d*q^2$ in Tesseract. Thus Tesseract allocates less memory to each processor than its predecessor.

In Figure \ref{fig:method}, the darkened part represents a single layer in Tesseract, each layer consists of $q * q$ processors, and $d$ numbers of layers construct the Tesseract structure. For each layer, it performs matrix multiplication individually, when comes to the need for synchronizing parameters, Tesseract will execute operations across the layers. Tesseract could be described as a further parallelized tensor parallelism with each layer's ability to perform matrix-multiplication operations and non-matrix-multiplication operations separately.

In this arrangement, processors could work on $d$ SUMMA-like matrix multiplications independently, thus reaching the target to reduce the computation time used. Compared to the 2.5-D matrix multiplication mentioned in section \ref{2.5}, our algorithm uses less memory on each processor and fewer transmissions between processors. Compared to the 2-D SUMMA algorithm, our work could make the matrices with different depths conduct matrix multiplication concurrently and relatively independently (except for necessary communication for parameter matrices), which could reduce the required time for matrix multiplication with a huge amount of data. As the lower bound of communication time and latency mentioned in 2.5-D algorithm, $W = \Omega(\frac{n^2}{\sqrt{dp}})$, $S =\Omega(\frac{p^{1/2}}{d^{3/2}})$, when $d > 1$, we have lower communication time and latency compared to 2-D algorithm. We could get the conclusion that with the same amount of processors, greater $d$ could lead to less communication and lower latency. In a special case $d = p^{1/3}$, we have $W = \Omega(\frac{n^2}{p^{2/3}}), S =\Omega(1)$, where the Tesseract could yield best efficiency.

Among all the parallelization structures, it is clear that the efficiency is negatively correlated with the number of processors, and positively correlated with the problem size assigned to each processor. We use isoefficiency function \cite{isoefficiency} to scale the efficiency of Tesseract. The parallel execution time of Tesseract is represented as 
\begin{equation}
    T_{para} = W/p + T_{comm},
\end{equation}
, where $W$ represents the serial execution time, $p$ represents the count of processes, $T_{comm}$ represents the time needed for communication. So the efficiency could be represented as 
\begin{equation}
    \mathrm{Efficiency} = \frac{W}{T_{para}p} = \frac{1}{1 + \frac{T_{comm}p}{W}}.
\end{equation}

For Megatron-LM, its $T_{comm}$ consists all-reduce operations, which makes its communication $\frac{2\beta (p-1)bsh}{p}$, and its isoefficiency function $W \sim p^3$, where $\beta$ denotes time to transfer a scalar. For Optimus, its $T_{comm}$ consists broadcast and reduce operations, which makes its communication time $\frac{2\beta bsh^2q\log p}{p}$, and its isoefficiency function $W \sim (\sqrt{p}\log p)^3$. For Tesseract, its $T_{comm}$ consists broadcast and reduce operations as well.

Tesseract reduces communication between GPU significantly as well, which is a huge drawback for Cannon's Algorithm and 2.5D matrix multiplication. With GPU amount $p$, Canon's Algorithm requires $2*p^{\frac{3}{2}}-2*p^{\frac{1}{2}}$ times of information transfer between GPU for a single matrix multiplication operation, 2.5D algorithm requires $2*p-2*p^{\frac{1}{3}}$ amount of transmission. For Tesseract, however, when $d = q$, requires only $2*p^{\frac{2}{3}}$ times transmission between GPU. In comparison, Tesseract requires less transmission with $q>2$ compared to Cannon's Algorithm, $q>4$ compared to the 2.5D algorithm. In the real situation, it usually requires more than four GPUs to parallelize the huge parameter, thus Tesseract requires less transmission between GPUs when training models.

\begin{algorithm}
   \caption{Tesseract matrix multiplication ($p$ processors in a $[q, q, d]$ shape)}
   \label{Tesseract algo}
\begin{algorithmic}
   \STATE {\bfseries Input:} Matrix $A$ with size $[a, b]$; Matrix $B$ with size $[b, c]$
   \STATE {\bfseries Output:} Matrix $C = A * B$ with size $[a, c]$
   \STATE split $A, B$ into partitioned matrices with shape of $[\frac{a}{qd}, \frac{b}{q}]$ and $[\frac{b}{q}, \frac{c}{q}]$ accordingly\;
   \STATE store $A_{ij}, B_{ij}$ into $p_{ij}$ accordingly\;
   \STATE store $C_{ij} = 0$ into $p_{ij}$ accordingly\;
   \FOR{$i, j$ {\bfseries in} $\{0,...,q-1\}$, $k$ {\bfseries in} $\{0,...,d-1\}$}
   \STATE store $B_{ij}$ into $p_{ijk}$\;
   \STATE $h = i + k * q$\;
   \STATE store $A_{hj}$ into $p_{ijk}$\;
   \STATE $C_{hj} = 0$\;
   \STATE store $C_{hj}$ into $p_{ijk}$\;
   \ENDFOR
   \FOR{$i, j$ {\bfseries in} $\{0,...,q-1\}$, $k$ {\bfseries in} $\{0,...,d-1\}$}
   \FOR{$t$ {\bfseries in} $\{0,...,q-1\}$}
   \STATE broadcast $A_{itk}$ in $p_{itk}$ to $p_{ijk}$\;
   \STATE broadcast $B_{tjk}$ in $p_{tjk}$ to $p_{ijk}$\;
   \STATE $C_{ijk} = C_{ijk} + A_{itk}*B_{vtk}$\;
   \ENDFOR
   \ENDFOR
   \STATE combine all $C_{ij}$ accordingly to $C$\;
   \STATE \Return $C$
\end{algorithmic}
\end{algorithm}

\subsection{Transformer on Tesseract}
\label{section:Transformer on Tesseract}
\begin{figure*}[!h]
  \centering
  \setbox1=\hbox{\includegraphics{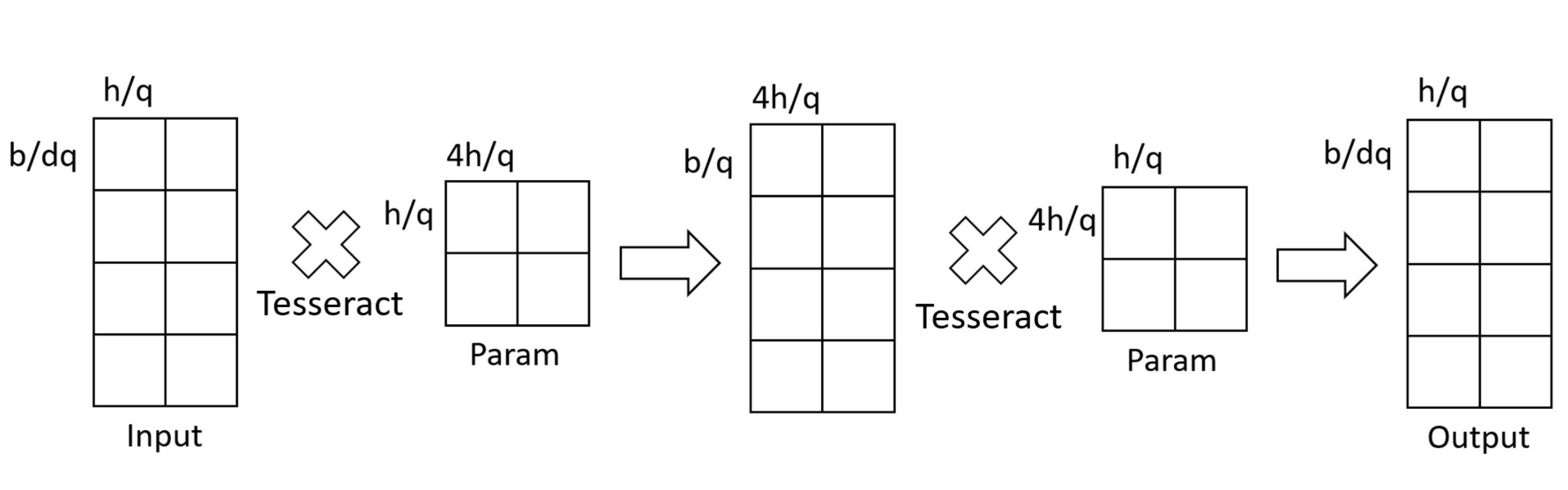}}
  \subcaptionbox{Feed forward\label{feedforward}}{\centering\includegraphics[width=125mm,scale=0.25]{figure/feedForward.pdf}}
  \qquad
  \setbox1=\hbox{\includegraphics{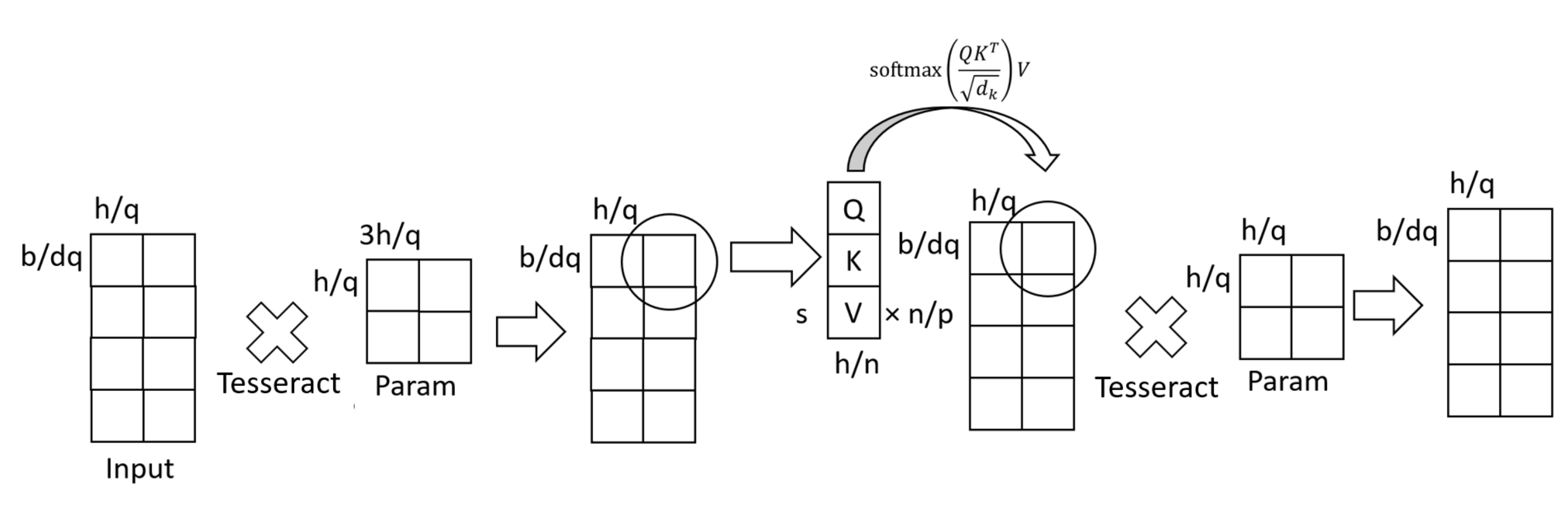}}
  \subcaptionbox{Multi-head attention\label{attention}}{\centering\includegraphics[width=125mm,scale=0.25]{figure/attention.pdf}}
  \caption{\label{Tesseract fig}Tesseract's procedure on both feed forward layer and multi-head attention layer, in this example, we used $q=2, d=2$ for presentation.}
\end{figure*}

In our work, we apply our Tesseract with Transformer as an example. There are encoders and decoders in Transformer, where the encoder consists of a multi-head attention layer and a feed forward layer (fully connected layer) with residual connection, and the decoder consists of two multi-head attention layers and a feed forward layer with residual connection. The main task for our work is to solve the multi-head attention layer and the feed forward layer since the residual connection is just to add vectors together and apply a normalization function. As shown in Figure \ref{Tesseract fig} we present both the multi-head attention layer and feed forward layer. In our work, there are two different types of operations, defined as matrix multiplication related and non-matrix-multiplication related, and they are treated with separated strategies to get the highest efficiency.
\\

\subsubsection{Matrix multiplication procedure}

\paragraph{Feed forward layer}

Feed forward layer could be understood as a simple multi-layer perceptron layer, and it will take an input matrix with a shape of $[b, s, h]$ by default, multiply two parts of parameters in the shapes of $[h, 4h]$ and $[4h, h]$ accordingly, and the output of feed forward layer would also be in the shape of $[b, s, h]$. In the Transformer, the purpose of using feed forward layer is to process the output matrix from one multi-head attention section to fit the next attention section as an input matrix. With the property of no communication with other input tokens nor inference issues, feed forward layer could be parallelized and thus improve the performance.

In our work, we split the input matrix into partitioned matrices with a shape of $[b/dq, s, h/q]$, and split the parameters into a shape of $[h/q, 4h/q]$ and a shape of $[4h/q, h/q]$ respectively. As shown in Figure \ref{feedforward}, our work applied Tesseract to parallelize the matrix multiplication. As the output, the parallelized feed forward layer will return partitioned matrices with a shape of $[b/dq, s, h/q]$ just as the input partitioned matrices.

\paragraph{Multi-head attention layer}

In this attention section, the input matrix with a shape of $[b, s, h]$ multiplies a $[h, 3h]$ weight matrix and get a $QKV$ matrix consisting of queries ($Q$), keys ($K$) and values ($V$) in the shape of $3 * [b, s, h]$. Then the $Q, K, V$ matrices will be partitioned into $n$ attention heads, the result matrices of $Q, K, V$ will be in shape $[s, h/n]$ for each sequence. Attention section gets an attention score $A$ in shape $[s, s]$ by performing $Q * K^T$, then gets the output of a single attention head by performing $A * V$. By gathering all the output of attention heads, the output will be in the shape of $[s, h]$. For all the sequences, the shape of the corresponding matrix is $[b, s, h]$, after the matrix multiplication with parameter matrix in shape $[h, h]$, the output shape will be $[b, s, h]$, just like the input shape. Similar to the feed forward section, with no communication with other position's tokens, the attention part is also parallelizable.

The procedure of our parallelized multi-head attention layer is shown in Figure \ref{attention}. In our implementation of the parallelized attention layer, we partitioned the input matrix into $[b/dq, s, h/q]$ matrices, with the partitioned $[h/q, 3h/q]$ parameter matrices, the resulted $Q, K, V$ matrices will be in a shape of $3 * [b/dq, s, h/q]$. There will be $n/q$ attention heads on each processor, and the received $Q, K, V$ matrix for each attention head has a shape of $[s, h/n]$. After the same procedure between $Q, K, V$ matrices, the output matrices will be combined together in the shape of $[b/dq, s, h/q]$.
After the matrix multiplication with $[h/q, h/q]$, the distributed processors will combine all the $[b/dq, s, h/q]$ matrices into the output $[b, s, h]$ matrix.

\subsubsection{Other procedure}

Besides feed forward and attention sections, there are sections in Transformer that are not suitable to apply parallelized operation, for example, the residual connection includes add and normalization operations. These kinds of sections will conduct operations locally on individual GPUs. For the bias-add operation, the matrices will be broadcast to each column for the forward process, and the backward process drives the gradients to be reduced back to the processor on row 0. As mentioned above, layer normalization is used in each residual connection, for better presentation, the result of layernorm function could be described as:
\begin{equation}
    \hat{X} = \frac{X - E[X]}{\sqrt{Var[X] + \epsilon}}.
\end{equation}
To compute the corresponding $E[X]=\frac{\Sigma X_i}{n}$, \\$Var[X]=E[X^2]-E[X]^2$, the processors will compute $X, X^2$ respectively and then run $all\_reduce$ function on each row. For the computation of gradient of $X$, the function could be described as:
\begin{equation}
    X^\prime = \frac{\frac{\delta J}{\delta \hat{X}}-\frac{(\Sigma \hat{X}_j(\frac{\delta J}{\delta \hat{X}})_j)\hat{X}+\Sigma (\frac{\delta J}{\delta \hat{X}})_j}{n}}{\sqrt{Var[X] + \epsilon}},
\end{equation}
for calculation, the processor will use stored $X, X^2$ and \\ $\frac{1}{\sqrt{Var[X] + \epsilon}}$, the calculation of will $\hat{X}_i(\frac{\delta J}{\delta \hat{X}}), \frac{\delta J}{\delta \hat{X}}$ take place similar to $X, X^2$. Due to the size of parameters, the communication loss in this process is negligible compare to the matrix multiplication part.

\subsection{Other models on Tesseract}
As mentioned in \ref{section:Transformer on Tesseract}, it is viable to implement Tesseract for models that is suitable for parallelization, for Tesseract is capable of both matrix-multiplication and non-matrix-multiplication procedures. With its adaptive solution on matrix multiplication, it could be used as the parallelization structure for many deep neural networks, for example, BERT\cite{BERT}, GPT-2\cite{gpt2}. With the ability to perform parallelized matrix multiplication and locally performed non-multiplication functions, Tesseract is able to perform different popular operations for neural networks, for example, residual modules, attention modules, and normalization modules. 

\begin{figure}[!h]
    \centering
    \includegraphics[width=80mm]{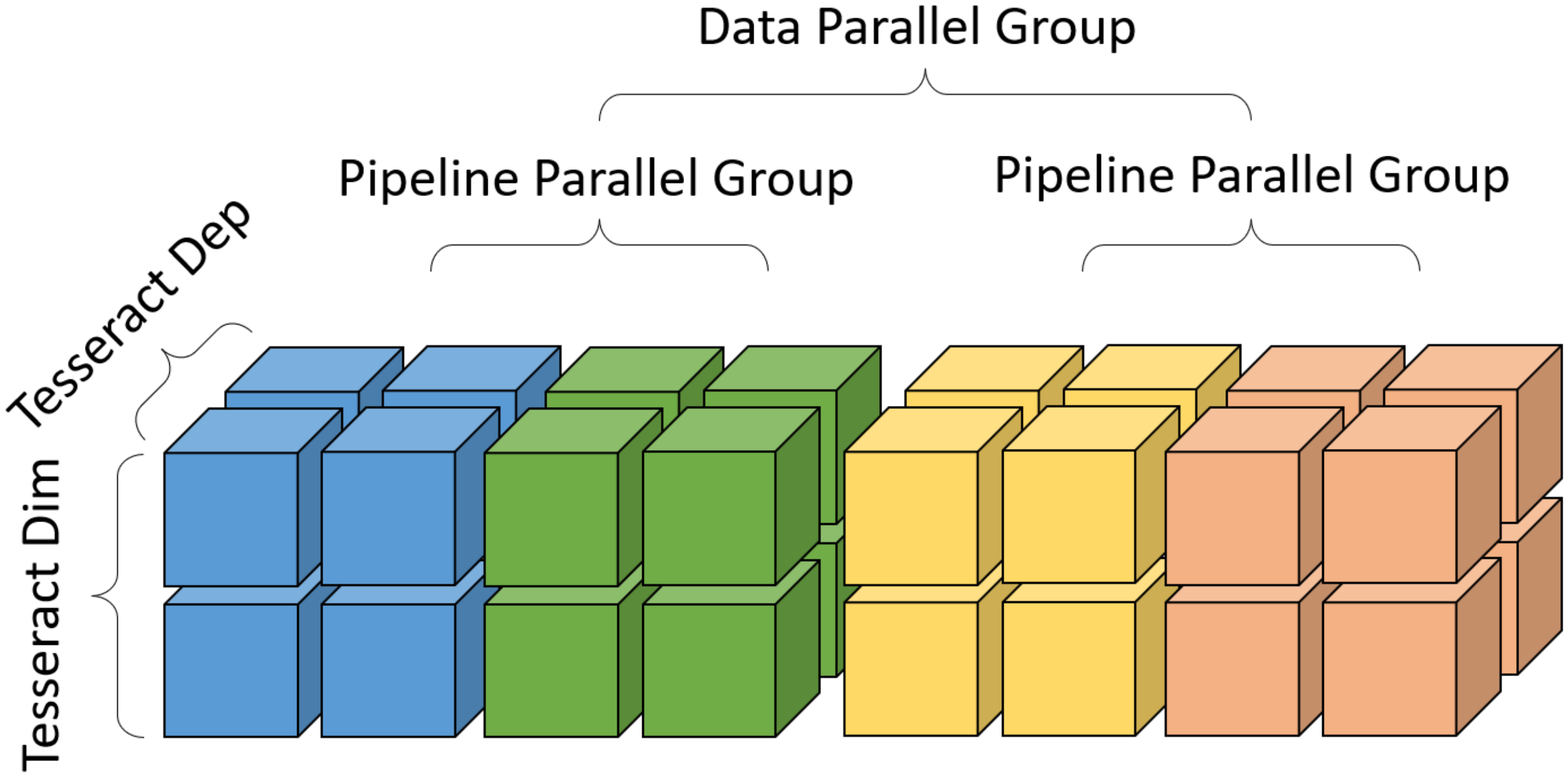}
    \caption{Structure of GPUs when applying Tesseract together with pipeline parallelism and data parallelism. Blocks in the same color represent GPUs in the same Tesseract module.}
    \label{fig:parallelism}
\end{figure}

\subsection{Compatibility with other model parallelisms}
Tesseract as a novel structure of tensor parallelism is capable to implement existing data parallelism and pipeline parallelism as well, to further enhance its performance. As shown in Fig \ref{fig:parallelism}, it demonstrates the distribution of GPUs with the data parallel size equals two, pipeline parallel size equals two, tesseract dimension equals two, and tesseract depth equals two. The number of total GPU involved will be 32 equals to data parallel size times pipeline parallel size times tesseract depth times square of tesseract dimension, where tesseract depth times square of tesseract dimension is tesseract size.

\section{Experiment}
\label{section:experiment}
Our experiments are conducted on the Meluxina server. On this supercomputer, there are 200 GPU nodes with 4 NVIDIA A-100 GPUs per node. Our experiments are conducted with 1, 2, 4, 8, and 16 nodes accordingly, and the number of GPUs ranging from 1 to 64, with settings: different Tesseract depths, fixed Tesseract dimension; different Tesseract dimensions and depths, fixed model parallel size. Strong scaling and weak scaling settings were applied to this experiment. Besides, all Tesseract experiment results are the averages of 20 times of the same experiments in this section.
We use randomly generated input matrices to check the algorithm and Xavier initialized parameter matrices. After the generation of matrices, we compute the matrix multiplication result and the result using our Tesseract method respectively, to guarantee outputs are the same. In this setting we could compare our Tesseract's time usage with that of other parallelized language models. 
In order to support Tesseract, we set different experiments compared with well-established 1-D parallelization structure Megatron-LM \cite{megatron} and 2-D parallelization Optimus \cite{optimus} in respective settings. 1-D and 2-D results are two strong baselines for Tesseract in our experiments. For the data listed below, we conducted our own experiments with the original codes of 1-D and 2-D acquired from GitHub.

For the interconnection between GPUs in our experiments, \\NVLink with a speed of 200 GB/s is used for communication within each node, and Infiniband with a speed of 200 Gbps is used for communication between nodes. Since communication cost between nodes is higher than communication within nodes, we arrange our experiments mainly by setting the size  $[q,q,d]$ where $q^2$ is a multiple of $4$. We set our experiments as above because Tesseract requires less communication between its $d$ layers.

In this part, $p$ represent the number of processors in each tensor parallel group, and we mark the shape of different parallelism with different representations:

\begin{itemize}
  \setlength\itemsep{1em}
  \item Tesseract shape: $[q, q, d]$ 
  \item Tesseract size: $p = q^2 * d$
  \item Megatron-LM shape: $[p]$ 
  \item Megatron-LM size: $p = p$
  \item Optimus shape: $[q, q]$ 
  \item Optimus size: $p = q^2$
\end{itemize}

\begin{table*}[!ht]
\caption{Comparison with different parallelization methods in strong scaling setting, the unit of measurement for forward time/batch and backward time/batch is second, and the unit of measurement for throughput and inference is the number of sequences per second.}
\begin{center}
\begin{tabular}{p{0.11\textwidth}p{0.05\textwidth}p{0.05\textwidth}p{0.05\textwidth}p{0.05\textwidth}p{0.05\textwidth}p{0.08\textwidth}p{0.08\textwidth}p{0.08\textwidth}p{0.08\textwidth}}
    \toprule
     parallelization & \#GPUs & GPU shape & batch size & hidden size & attention heads & forward time/batch & backward time/batch & throughput & inference\\
    \hline
    \multirow{3}{*}{Megatron-LM}&4&[4]&12&3072&64&0.1225&0.4749& 1.6739 & 8.1633\\
    &16&[16]&12&3072&64&0.1143&0.4293& 1.8396 & 8.7489\\
    &64&[64]&12&3072&64&0.1195&0.5306& 1.5382 & 8.3682\\
    \hline
    \multirow{3}{*}{Optimus}&4&[2,2]&12&3072&64&0.1676&0.5019& 1.4937 & 5.9666\\
    &16&[4,4]&12&3072&64&0.2099&0.6159& 1.2109 & 4.7642\\
    &64&[8,8]&12&3072&64&0.1329&0.3986& 1.8815 & 7.5245\\
    \hline
    \multirow{3}{*}{Tesseract}
    &4&[2,2,1]&12&3072&64&0.1666&0.5014& 1.4970 & 6.0024\\
    &8&[2,2,2]&12&3072&64&0.0999&0.3002& 2.4994 & 10.0100\\
    &16&[4,4,1]&12&3072&64&0.1444&0.4343& 1.7280 & 6.9252\\
    &32&[4,4,2]&12&3072&64&0.1244&0.3727& 2.0117 & 8.0386\\
    &64&[4,4,4]&16&3072&64&\textbf{0.0869}&\textbf{0.2636}& \textbf{2.8531} & \textbf{11.5075}\\
    &64&[8,8,1]&12&3072&64&0.1799&0.5178& 1.4333 & 5.5586\\
    \bottomrule
\end{tabular}
\end{center}
\label{strong-scaling}
\end{table*}

\begin{table*}[!ht]
\caption{Comparison with different parallelization methods in weak scaling setting, the unit of measurement for forward time/batch and backward time/batch is second, and the unit of measurement for throughput and inference is the number of sequences per second.}
\begin{center}
\begin{tabular}{p{0.11\textwidth}p{0.05\textwidth}p{0.05\textwidth}p{0.05\textwidth}p{0.05\textwidth}p{0.05\textwidth}p{0.08\textwidth}p{0.08\textwidth}p{0.08\textwidth}p{0.08\textwidth}}
    \toprule
     parallelization & \#GPUs & GPU shape & batch size & hidden size & attention heads & forward time/batch & backward time/batch & throughput & inference\\
    \hline
    \multirow{3}{*}{Megatron-LM}&4&[4]&60&2048&32&0.0793&0.2613&2.9360 & 12.6103\\
    &16&[16]&60&4096&64&0.2081&0.5149&1.3831 & 4.8054\\
    &64&[64]&30&8192&128&0.4638&1.0963&0.6410 & 2.1561\\
    \hline
    \multirow{3}{*}{Optimus}&4&[2,2]&96&2048&32&0.0827&0.2445&3.0562 & 12.0919 \\
    &16&[4,4]&192&4096&64&0.1829&0.5458& 1.3723 & 5.4675\\
    &64&[8,8]&384&8192&128&0.1962&0.5964& 1.2617 & 5.0968\\
    \hline
    \multirow{3}{*}{Tesseract}&1&[1,1,1]&48&1024&16&0.0603&0.1669& 4.4014 & 16.5837\\
    &4&[2,2,1]&96&2048&32&0.0867&0.2557& 2.9206 & 11.5340\\
    &8&[2,2,2]&192&2048&32&0.0864&0.2552& 2.9274 & 11.5741\\
    &16&[4,4,1]&192&4096&64&0.1177&0.3553& 2.1142 & 8.4962\\
    &32&[4,4,2]&384&4096&64&0.1173&0.3521& 2.1304 & 8.5251\\
    &64&[4,4,4]&768&4096&64&0.1155&0.3468& 2.1631 & 8.6580\\
    &64&[8,8,1]&384&8192&128&0.1799&0.5178& 1.4333 & 5.5586\\
    \bottomrule
\end{tabular}
\end{center}
\label{weak-scaling}
\end{table*}

\subsection{Strong scaling}
In the strong scaling setting experiments, as shown in Table \ref{strong-scaling}, we fix the problem size. The batch size is set as 12, whereas in the experiment with a Tesseract shape [4,4,4], the batch size is 16 since the batch size needed to be divisible by the product of Tesseract dimension and Tesseract depth $d*q$, and this change does not affect the result significantly, if any, only making the result worse than the actual result. The hidden size of the transformer is 3072 and the number of attention heads in strong scaling experiments is fixed at 64, due to the limitation of memory on each processor. 

From the results of the experiments, it manifests that with the same Tesseract dimension, greater Tesseract depths will reduce the forward/backward time per batch size, but not in inverse proportion since the communication loss will also increase. Besides, from the example between [4,4,4] and [8,8,1] arrangements, with the same number of processors, greater Tesseract depths could lead to less forward/backward time.

In the first situation, due to the increment of communication cost, double the depth could not result in a half forward/backward time as expected. While the outcome of the experiment in strong scaling between [4,4,4] arrangement and [8,8,1] arrangement could be because of the high communication loss, as the latter arrangement is supposed to have smaller matrices to compute. According to Table \ref{strong-scaling}, from the results of Tesseract with $q = 4$, we could find out that compared with Tesseract setting with a depth of 1, Tesseract with a depth larger than 1 could outperform by a large margin. Besides, when using the same amount of GPU, for example, 64, our Tesseract in the shape of [4,4,4] has a huge speedup compared with a 2-D solution with a shape of [8,8,1] calculated by $0.1799 / 0.0869 = 2.0702$, this could help to prove that with bigger depth at same amount processors, the distributed language model will have higher efficiency. 

Besides the comparison between the Tesseract with different depth settings, we can also calculate the speedup of Tesseract between Megatron-LM and Optimus. Here we focus on the experiments with 64 GPUs in different parallelization structures, compared to Megatron-LM, our Tesseract reached a speedup of $0.1195 / 0.0869 = 1.3751$, compared to Optimus, Tesseract reached a speedup of $0.1329 / 0.0869 = 1.5293$. With this huge speedup and the highly scalable feature of Tesseract, it will make the training of deep neural network models faster and more efficient.

\subsection{Weak scaling}
In the weak scaling setting experiments, the parameter setting depends on the individual arrangement of GPUs, in order to allocate the same problem size on each GPU. As shown in Table \ref{weak-scaling}, for all three different structures, we set the corresponding hidden size and number of attention heads to get a fixed size of input parameters, and the setting conforms $[b/dq, n/q, h/n] = [24, 16, 192]$(change of $n$ does not affect the result) due to the memory of each GPU used. 
From the comparison between settings of $[4,4,4], [8,8,1]$ with the same amount of total processors, we could come up with the enhancement of efficiency by computing $0.1799 / 0.1155 = 1.5576$, this proves that with bigger depth at the same amount processors, the distributed language model will have higher efficiency.

Throughput is calculated as the ratio of batch size to the sum of forward and backward time per iteration, and inference is the ratio of batch size to the sum of forward time per iteration only. For this weak scaling setting experiment, the memory used for each GPU is the same, but the communications required between GPUs are different, thus the throughput and inference speed are naturally faster for experiments with fewer GPUs used. For this reason, we compared the results using the same amount of GPUs instead. From our experiments' results, it shows that Tesseract reaches a maximum of $2.1631 / 0.6410 = 3.3746$ times the throughput compared to Megatron-LM and $2.1631 / 1.2617 = 1.7144$ times the throughput compared to Optimus in the setting of 64 total GPUs. For inference, Tesseract reached a maximum of $8.6580 / 2.1561 = 4.0156$ times the inference compared to Megatron-LM and $8.6580 / 5.0968 = 1.6987$ times the inference compared to Optimus. This result support that our Tesseract provides better utilization of GPU server compared to other tensor parallelisms 1-D and 2-D.

For the comparison between Tesseract's $[4,4,4]$ and $[8,8,1]$ settings which both use 64 GPUs in total, we find that by offering a bigger depth on Tesseract, the communication required for Tesseract is reduced significantly, for throughput, $[4,4,4]$ achieves $2.1631 / 1.4333 = 1.5092$ times compared to $[8,8,1]$, and \\$8.6580 / 5.5586 = 1.5576$ times inference.

\subsection{Performance on Neural Network Training}
\begin{figure}[!h]
    \centering
    \includegraphics[width=80mm]{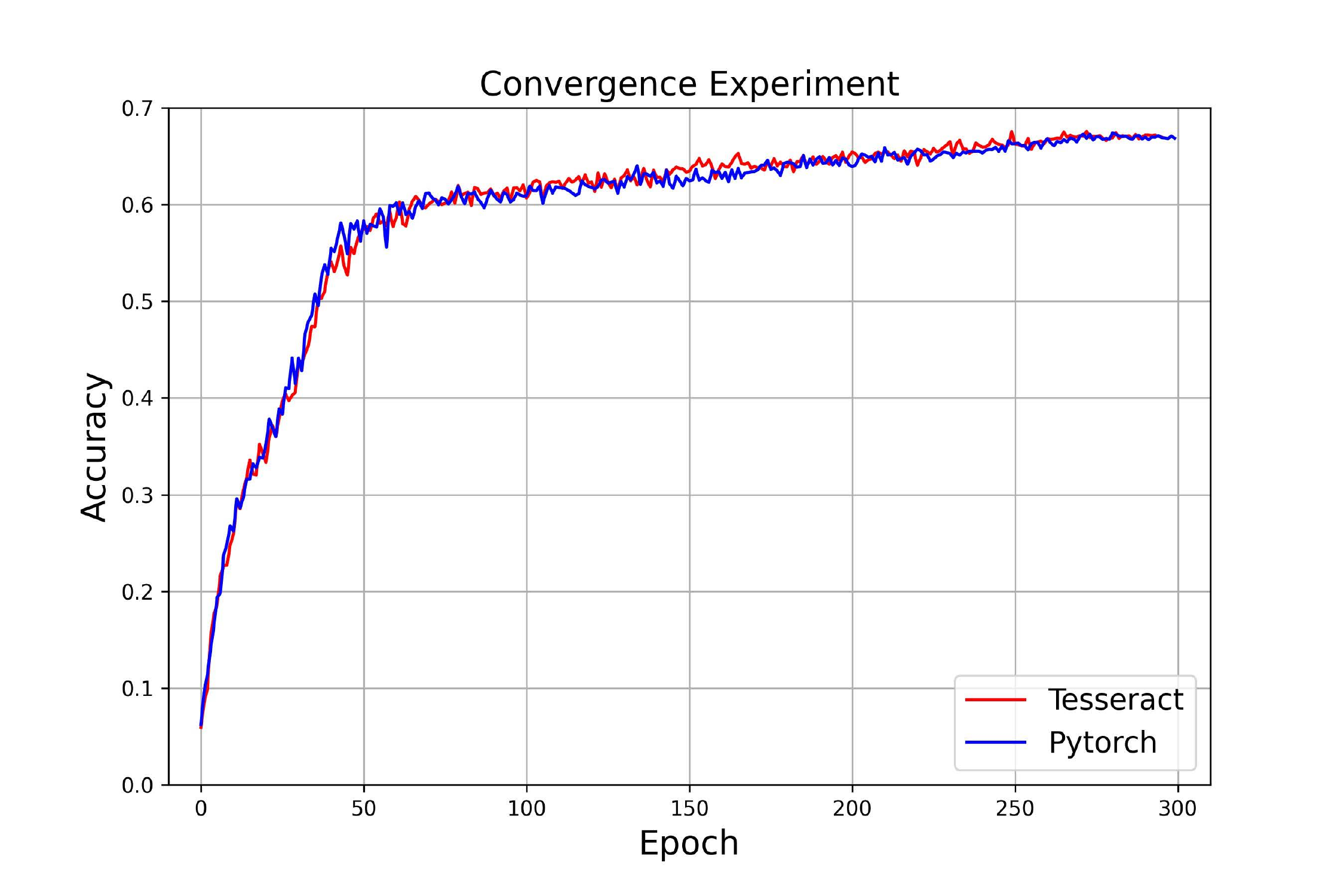}
    \caption{Training accuracy for ImageNet-100 dataset using Vision Transformer, in settings of (1) single GPU (2) Tesseract with shape [2,2,1] (3) Tesseract with shape [2,2,2]. This experiment includes (1) as the baseline, (2), and (3) represents different settings of tesseract depth.}
    \label{fig:convergence}
\end{figure}

Tesseract does not introduce any approximations, thus it does not affect the training accuracy for neural networks. We conducted experiments using Tesseract on Vision Transformer\cite{dosovitskiy2020vit} to discuss our performance on accuracy and throughput. As shown in Figure \ref{fig:convergence}, we compared the accuracy results of a Vision Transformer model with the same parameters on different GPU settings. In this Vision Transformer model, the number of epochs is 300, batch size is 512, Adam is used as the optimizer, and the learning rate is 0.003 with a weight decay of 0.3. To control the variables, we fixed random seeds and initialization methods for the experiment. The training dataset is ImageNet-100\cite{ImageNet}. According to our accuracy results, Tesseract does not affect the model's accuracy.

\section{Conclusion}
\label{section:conclusion}
In our work, we design a further parallelized tensor parallelism structure - Tesseract, and we tested this work on the computer vision model Vision Transformer\cite{dosovitskiy2020vit} and language model \\Transformer\cite{Transformer}. By using Tesseract, we split the input matrices and parameter matrices according to the shape of the arrangement of processors in the group, thus reducing the time used for calculation. Besides, compared to previous tensor parallelisms, our Tesseract requires less memory on each GPU and it requires less transmission of data among GPUs, which in return could lead to higher efficiency and less communication loss. In our evaluation, Tesseract outperforms both 1-D and 2-D parallelization methods, and we also evaluated the impact of the setting of the 'depth' parameter in this structure, the conclusion is with the same total amount of processors, greater depths could further increase the efficiency of Tesseract. Moreover, we have tested the training accuracy of frequently used deep neural network models. We study the effect of Tesseract on the training accuracy of ImageNet with Vision Transformer and conclude Tesseract does not harm the convergence of neural networks. 

\bibliography{citations}
\bibliographystyle{ACM-Reference-Format}

\end{document}